\documentstyle[epsf,graphicx,aps,prl,multicol,german]{revtex}

\begin{document}
\draft

\wideabs{
\title{Origin of Anomalous Diffusion and Non-Gaussian Effects for Hard Spheres: Analysis of Three-Time Correlations}
\date{September 1998}
\author{B. Doliwa and A. Heuer}
\address{Max-Planck-Institut f\"ur Polymerforschung, Postfach 3148\\ D-55021 Mainz, Germany}
\maketitle

\begin{abstract}
We present new simulation results on a hard sphere system at high densities.
Using three-time correlations, we can account for the anomalous diffusion, which results from a
homogeneous back-dragging effect.
Furthermore, we calculate the non-gaussian parameter and connect it to the existence of
dynamic heterogeneities.
\end{abstract}
}

\newcommand{\ppara}{p_{\|}(x_{12}|r_{01}; t)}
\newcommand{\psenk}{p_{\perp}(y_{12}|r_{01}; t)}
\newcommand{\pparat}{p_{\|}(x_{12}|r_{01})}
\newcommand{\psenkt}{p_{\perp}(y_{12}|r_{01})}
\newcommand{\pparao}{p_{\|}(\vec r_{12}\cdot\widehat{r}_{01}|r_{01}; t)}
\newcommand{\psenko}{p_{\perp}(\vec r_{12}\cdot\widehat{u}_{01}|r_{01}; t)}
\newcommand{\pzz}{p_{zz}(z_{12}|z_{01}; t)}
\newcommand{\pzza}{p_{zz}(z_{12}| |z_{01}|; t)}
\newcommand{\pzzat}{p_{zz}(z_{12}| |z_{01}|)}
\newcommand{\vra}{\vec r_{01}}
\newcommand{\bra}{\widehat r_{01}}
\newcommand{\vrb}{\vec r_{12}}
\newcommand{\brb}{\widehat r_{12}}
\newcommand{\vua}{\vec u_{01}}
\newcommand{\bua}{\widehat u_{01}}
\newcommand{\vub}{\vec u_{12}}
\newcommand{\bub}{\widehat u_{12}}
\newcommand{\cs}{c_\perp(r_{01})}
\newcommand{\cp}{c_\|(r_{01})}
\newcommand{\czz}{c_{zz}(z_{01})}
\newcommand{\ceff}{c_{\textrm{\footnotesize eff}}(t)}
\newcommand{\ceffq}{c^2_{\textrm{\footnotesize eff}}(t)}
\newcommand{\xb}{x_{12}}
\newcommand{\yb}{y_{12}}
\newcommand{\ra}{r_{01}}
\newcommand{\za}{z_{01}}
\newcommand{\zb}{z_{12}}
\newcommand{\sizz}{\sigma_{zz}(z_{01})}
\newcommand{\sip}{\sigma_\|(r_{01})}
\newcommand{\sis}{\sigma_\perp(r_{01})}
\newcommand{\at}{\alpha_2(t)}
\newcommand{\ato}{\alpha_2}
\newcommand{\ai}{\alpha_2(\infty)}
\newcommand{\alp}{\alpha_{2,\textrm{\footnotesize poly}}(t)}
\newcommand{\apo}{\alpha_{2,\textrm{\footnotesize poly}}}
\newcommand{\api}{\alpha_{2,\textrm{\footnotesize poly}}(\infty)}
\newcommand{\Ta}{t_{\alpha_2}}
\newcommand{\ta}{\tau_\alpha}
\newcommand{\am}{\alpha_{2,\textrm{\footnotesize max}}}
\newcommand{\sqrr}{\langle r^2(t)\rangle}
\newcommand{\zza}{\langle z_{01}^2\rangle}
\newcommand{\zzb}{\langle z_{12}^2\rangle}
\newcommand{\zzc}{\langle z_{02}^2\rangle}
\newcommand{\zzab}{\langle z_{01}z_{12}\rangle}

\section{Introduction}

In recent years the nature of the non-exponential relaxation of glass-forming liquids close to $T_g$
has been elucidated by a variety of experimental techniques; see \cite{Ediger} and \cite{Boehmer} for
an overview.
One way of getting additional information is to analyse three-time besides the standard two-time
correlation functions, as realized by multidimensional NMR experiments.
On this basis it is possible to decide whether the non-exponentiality is mainly due to a homogeneous scenario
(intrinsic non-exponentiality related to systematic back- and forth dynamics)
or a heterogeneous scenario (superposition of different exponential processes).
In a previous publication these concepts have been applied to computer simulations of a hard sphere system
which experimentally is well represented by colloidal suspensions \cite{vanMegen}.
There we observed that the dynamics at short times ($\beta$-regime) is dominated
by the presence of a cage formed by surrounding particles.
This results in a systematic back-dragging effect, corresponding to a mainly homogeneous scenario.
In the $\alpha$-regime
most particles have escaped their initial cages. Then they encounter fast or slow regions in the liquid,
arising, e.g., from inhomogeneitities in density. 
The simplest model for
this situation would be an ensemble of independently diffusing particles with different mobilities.

The goal of the present work is to quantify the hard sphere dynamics in terms of
homogeneous and heterogeneous contributions for all time scales.
In extension of our previous work, we introduce appropriate measures for homogeneous and heterogeneous contributions.
They help to elucidate the physics of the sublinear diffusion in the $\beta$-regime and the non-gaussian
parameter.

\section{Simulation Details}
We perform conventional Monte Carlo dynamics with periodic boundary
conditions on a system of a thousand hard spheres with a size polydispersity of 10 percent.
The algorithm is described in \cite{Cichocki,PRL}. 
For the $\beta$ and $\alpha$ regimes, the mean step size has been chosen to give an
acceptance rate of 50 percent for each move. We checked, that, apart from an overall scaling,
the dynamics is not influenced by changing the mean step size for these time regions.

At very short times, we perform separate simulation runs with a reduced step size, in order to obtain the
microscopic dynamics. This is necessary, because we want to define a density-independent time scale by
fixing the short time diffusion constant, i.e. $\sqrr=6D_0t$, where t is very small.
We use $D_0\equiv\frac1{160}$ and set the unit of length equal to the mean particle diameter,
as is done by Fuchs et al. \cite{Fuchs}.

The highest density, analysed in this paper, is $\phi\equiv\langle\frac43\pi R^3_i\rangle \frac NV=0.58$.
At this high packing fraction, we have chosen a very long equilibration time ($5\cdot 10^7$ MCS) to avoid
aging effects during the measurement.
Interestingly, the non-gaussian parameter (NGP) turns out to be a very sensitive indicator for the degree of equilibration.

\begin{figure}[t]
\centerline{\includegraphics[clip, width=3.25in]{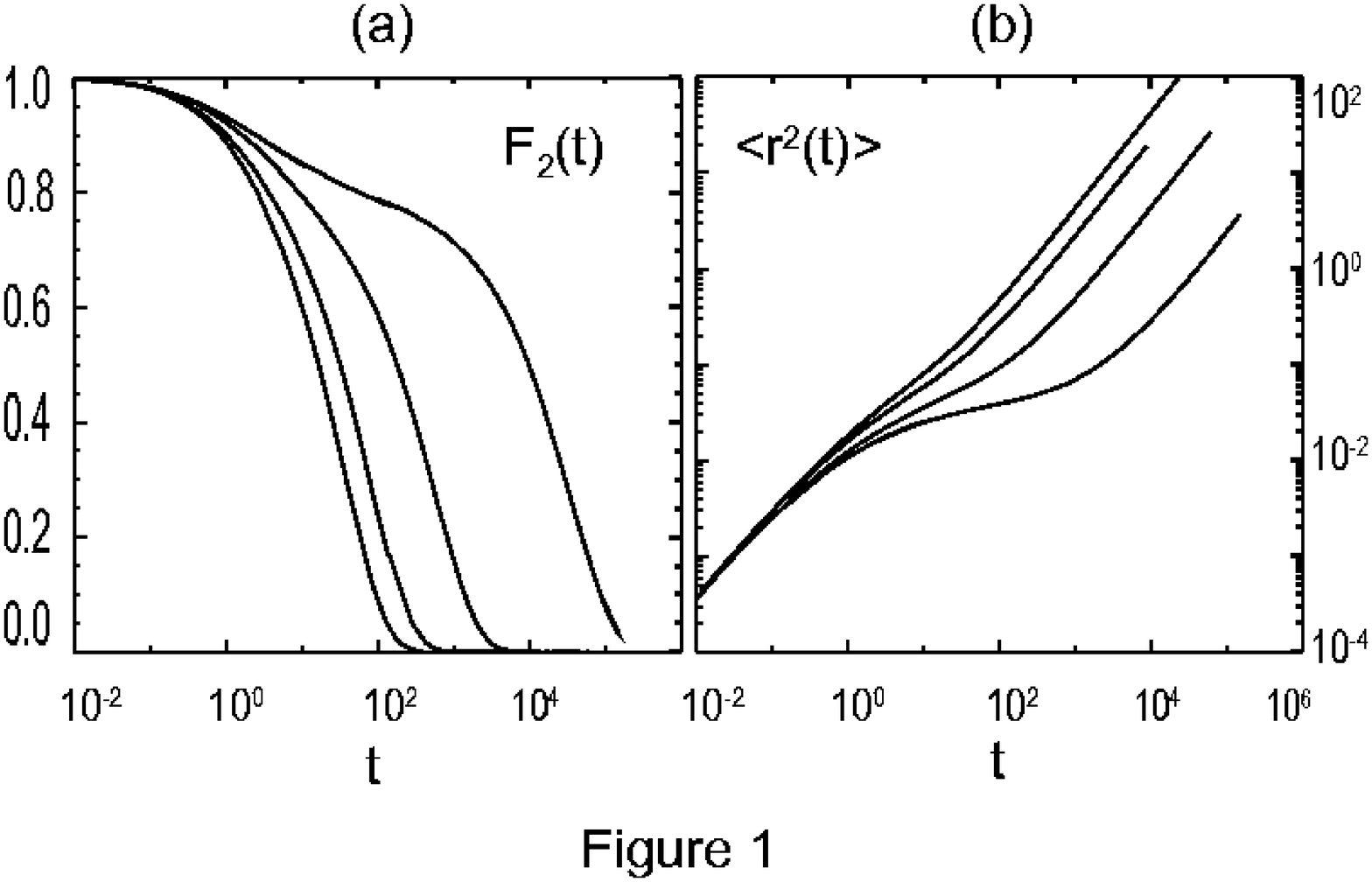}}
\renewcommand{\figurename}{Fig.}
{\footnotesize
(a) The two-time correlation function $F_2(t)=\langle\cos \vec q(\vec r(t)-\vec r(0))\rangle$ for $q=2\pi$
which is close to the maximum of the structure factor. The curves are plotted for the densities 
$\phi=50, 53, 56,$ and $58 \%$, from left to right.
(b) The mean square displacements $\sqrr$ for the densities in (a), again from left to right.
}
\label{fig:scattANDsqrr}
\end{figure}

In figure \ref{fig:scattANDsqrr} we plot the incoherent scattering function
$F_2(t)=\langle\cos \vec q(\vec r(t)-\vec r(0))\rangle$ for $q=2\pi$ and the mean square displacement,
clarifying the time scales of our simulations.

\section{Three-Time Correlations: The Homogeneous Part}
As shown in our recent work \cite{PRL}, we can gain new insight into the dynamics by analyzing
three-time correlation functions. For a given time $t$ we define the conditional probabilities 
$\pparao$, $\psenko$ and $\pzz$ with
$\vec r_{mn}\equiv\vec r(nt)-\vec r(mt)$, $r_{mn}\equiv|\vec r_{mn}|$.
Here $\bua$ is an arbitrary vector perpendicular to
$\vra$, the hat denoting a unit vector. We abbreviate
$x_{12}\equiv\vrb\cdot\bra$ and
$y_{12}\equiv\vrb\cdot\bua$ for later purposes. Thus, $\ppara$ denotes the probability, that a particle moves
during the second time interval
the distance $x_{12}$, projected onto the vector $\bra$, under the condition that it made a distance $\ra$ in
the first time interval. The probability function $\psenk$ is defined analogously, yielding
information about the distance $y_{12}$ travelled perpendicular to the first step.
Finally, $\pzz$ includes the information about the projection of both steps onto a randomly chosen direction.

\begin{figure}[t]
\centerline{\includegraphics[width=3in, clip]{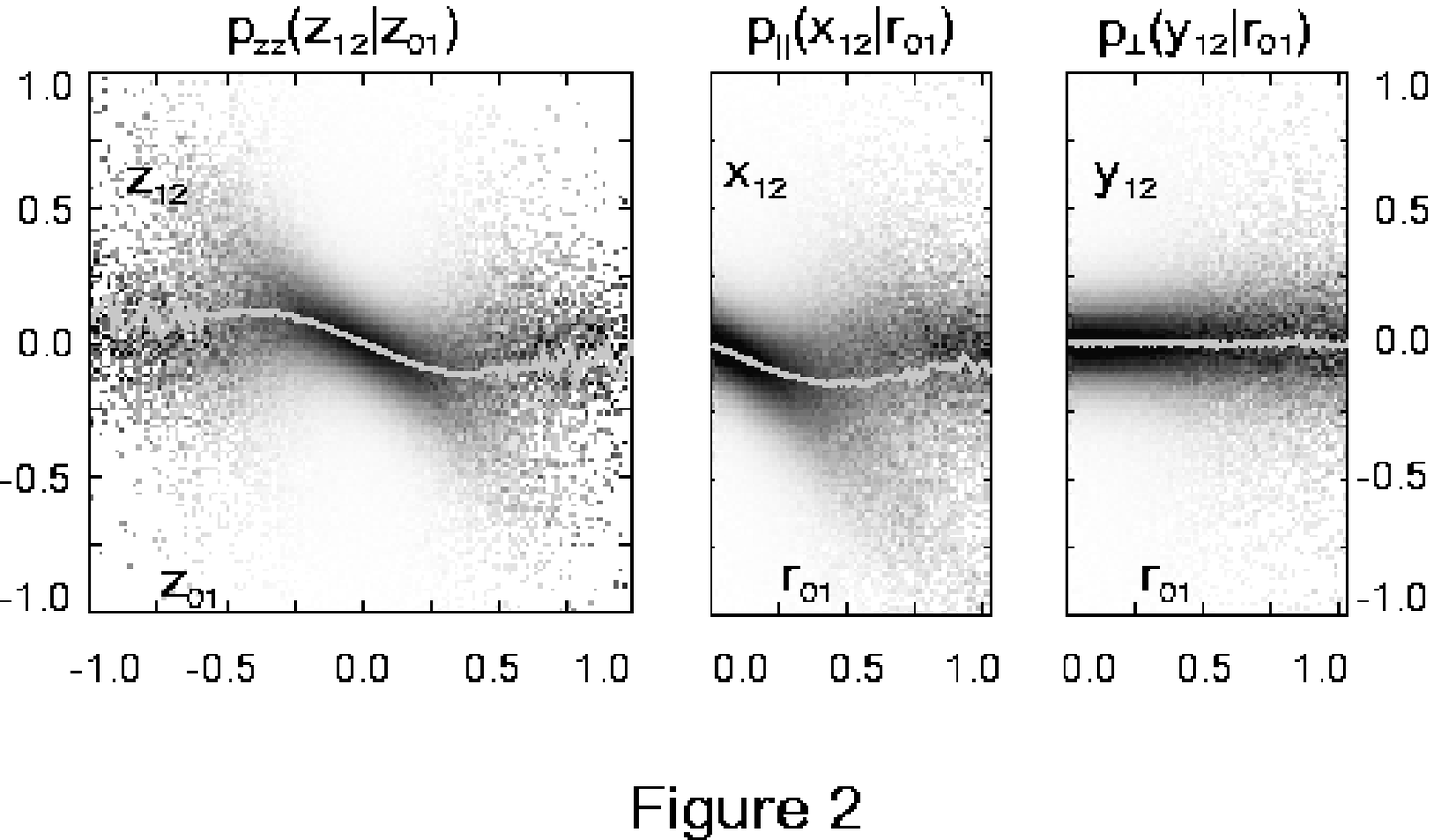}}
\renewcommand{\figurename}{Fig.}
{\footnotesize
The conditional probabilities $\pzz$, $\ppara$,
and $\psenk$ for a time $t\approx400$ in the $\beta$-regime,
where $x_{12}\equiv\vrb\bra$ and $y_{12}\equiv\vrb\bua$.
$\za$ and $\zb$ denote the projections of the subsequent displacements onto an arbitrary direction.
The dark areas correspond to high probabilities.
The average values $\czz$, $\cp$, and $\cs$ are indicated by the light lines.
Note that the unit of length is greater than in \cite{PRL} by a factor of 2.
}
\label{fig:allall}
\end{figure}

For non-exponential relaxation, as displayed by supercooled liquids, memory effects are present, resulting
in a strong dependence of $p_{zz}$ and $p_{\|,\perp}$ on $\za$ and $\ra$, respectively.
We see the typical situation in figure \ref{fig:allall}, where
all these probability distributions are shown for a density of $\phi=58\%$ at a time $t\approx 400$ in the $\beta$-region.
The main features of these plots are the non-zero displacements,
$\czz\equiv\langle \zb\rangle$,
$\cp\equiv\langle x_{12}\rangle$ and
$\cs\equiv\langle y_{12}\rangle$ as indicated,
and the widths
$\sizz$,
$\sip$ and
$\sis$, where, e.g.,
$$\sigma^2_\|(r_{01})\equiv\int dx_{12}\pparat (x_{12}-\cp)^2.$$
The possible dependence on t of these quantities should be kept in mind.

As expected from the definition of $p_\perp$, the value of $\cs$ must be zero.
We discussed in \cite{PRL} that in the heterogeneous limit, there is no back-dragging effect, i.e.
$\czz=\cp=0$ whereas in the homogeneous case we have $\sigma_{zz,\|,\perp}(\ra)=const$ and $c_{zz,\|}(\ra)=const.$

As shown in \cite{PRL}, $\czz$
- which, by the way, is equal to $\cp$ -
can account for anomalous diffusion in the $\beta$-domain.
To be precise, we made use of the fact that $\czz\approx -c\za$, where $c=c(t)$ is a constant.
If we additionally assume that $p_{zz}(\cdot|\za)$ is gaussian with constant width $\sizz=\sigma$,
straightforward integration yields a connection
between $c(t)$ and the approximate logarithmic slope
$$\beta(t)\equiv\frac{\ln \langle r^2(2t)\rangle -\ln \sqrr}{\ln 2t - \ln t}
\approx\frac d{d \ln t}\ln\sqrr$$
of the mean square displacement, namely,
\begin{eqnarray}
\beta(t)\approx1+\frac{\ln(1-c(t))}{\ln2}.
\end{eqnarray}

This estimate becomes wrong for longer times, because the back-dragging is not linear anymore.
Mainly this comes from the fact that particles start to leave their cages,
resulting in a more complex behaviour.
This may explain the deformation of $\czz$ and $\cp$, which now look more like the first part of a
negative sine. To account for this effect, we define an 'effective' slope of $\czz$, namely,
\begin{eqnarray}
\ceff\equiv-\frac{\int {\textrm d}\za p(\za)\za\czz}{\langle \za^2\rangle},
\end{eqnarray}
where all quantities depend on t.

Using $\ceff$ instead of $c(t)$ as input for Eq. (1), 
the approximation of $\beta$ becomes exact. 
With $\sqrr=3\zza=3\zzb$, we find
\begin{eqnarray}
	\	\zzc = \langle (z_{01}+z_{12})^2 \rangle = 2(\zza+\zzab) \nonumber\\
\Leftrightarrow \frac\zzc\zza = 2(1+\frac\zzab\zza)  \nonumber \\
\Leftrightarrow  \beta \equiv\frac{\ln\zzc-\ln\zza}{\ln2} = 1+\frac{\ln(1+\frac\zzab\zza)}{\ln2} \nonumber \\
		=1+\frac{\ln(1-\ceff)}{\ln2} \nonumber.
\end{eqnarray}

If we had the purely heterogeneous scenario, where the conditional probability for the second move
does only depend on the length of the first one, i.e. $\pzz=\pzza$, then
$$\zza\ceff=\int {\textrm d}\za{\textrm d}\zb \pzzat p(\za) \za\zb =0,$$
because $p(\za)$ is an even function.
Therefore, $\ceff$ can be regarded as an appropriate measure for homogeneous contributions, and
in the linear case, it reduces to the slope $c(t)$.
Hence homogeneous contributions strongly influence the nature of the anomalous diffusion.

\section{The Non-Gaussian Parameter}

A way to approach the high-density features of the self part of the van Hove correlation function is to
analyze the non-gaussian parameter
$$ \alpha_2(t)\equiv \frac{3\langle r^4(t)\rangle}{5\langle r^2(t)\rangle^2}-1.$$
It measures the deviation from gaussian behaviour and consequently has to vanish for $t\to 0$, 
because the dynamics on the very microscopic scale is brownian by definition.

The NGP has been calculated for other systems, e.g. Lennard-Jones liquids \cite{LJgross}
or soft disks \cite{softdisks}. In every case, a maximum has been found in the $\beta$ region,
which is the more pronounced, the lower the temperature is.

\begin{figure}[t]
\centerline{\includegraphics[clip, width=3.25in]{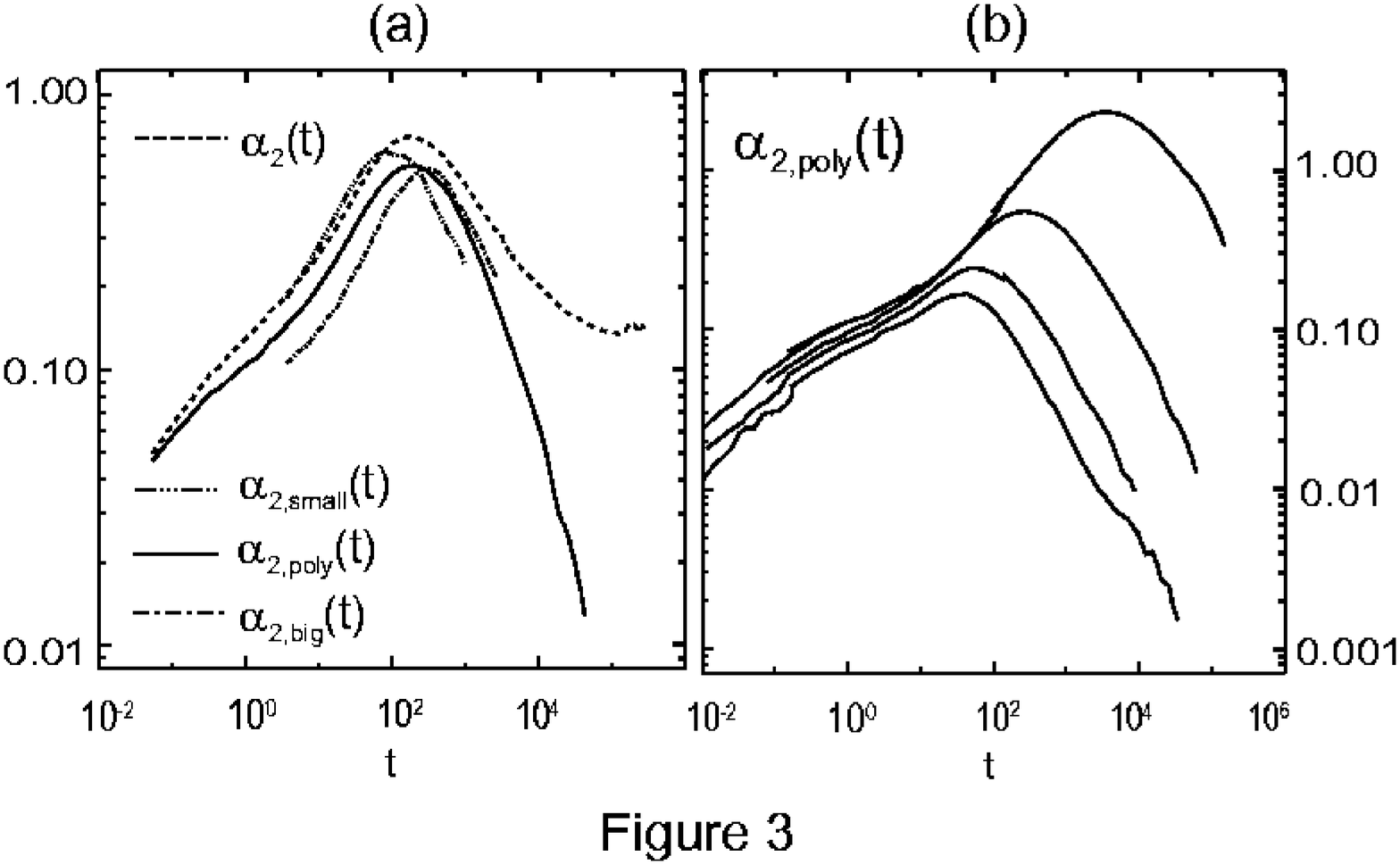}}
\renewcommand{\figurename}{Fig.}
{\footnotesize
(a) The non-gaussian parameter $\alpha_2\equiv(3\langle r^4\rangle)/(5\langle r^2\rangle^2)-1 $
for density $\phi=56 \%$, calculated in different ways.
For $\alpha_{2,\textrm{\footnotesize small}}$, $\alpha_{2,\textrm{\footnotesize big}}$ and $\alpha_2$
the average is over the $15 \%$ smallest, the $15 \%$ biggest and all particles, respectively.
In $\alpha_{2,\textrm{\footnotesize poly}}$ the different sizes are treated separately, averaging
afterwards, as discussed in the text.
(b) The NGP $\alpha_{2,\textrm{\footnotesize poly}}$ for the densities
$\phi=50, 53, 56,$ and $58 \%$, from bottom to top.
Notice that the axes in (a) and (b) have been chosen logarithmically.
}
\label{fig:compNGP_56}
\end{figure}

In figure \ref{fig:compNGP_56} (a) we see $\at$ for $\phi=56\% $ as a dashed line.
It clearly does not decay to zero, but assumes a value $\ai\approx 0.15$.
However, we must be careful with the definition of $\ato$, because the polydispersity
can cause a trivial non-gaussianity, due to a size-dependent particle mobility.
Therefore, we should calculate the NGP for every particle diameter separately, and take the mean value
over the size distribution afterwards, i.e.
$$ \alp\equiv\langle\alpha_{2,R}(t)\rangle_{R}.$$
Figure \ref{fig:compNGP_56} (a) shows this quantity for comparison.
Thus, at this density, $\api=0.$ 
The other two curves show the NGP, averaged over the $15 \%$ largest and smallest particles, respectively.
Differing in their maximum values by about $10 \%$, they exhibit a time separation of their maxima by a factor of three.
Kob et al. have found the same dependence for their binary Lennard-Jones mixture \cite{LJgross}.

In figure \ref{fig:compNGP_56} (b), we see $\alp$ for different volume fractions in a double logarithmic plot.
For small times, one perceives a linear ascend which corresponds to an exponent of approximately $0.3$.
Later on, $\apo$ reaches its maximum value at a time $\Ta$ in the late $\beta$-region and then slowly
decreases again.
As we can see, the maximum value of $\apo$ grows with density,
i.e. $\am=0.17,0.24,0.56,2.33$, for $\phi=0.50,0.53,0.56,0.58$, respectively. This strong dependence indicates a
tremendous change of the dynamics for densities $\phi>0.56$.

\section{Three-Time Correlations: The Heterogeneous Part}
 
It has been suggested that the value of $\at$ is intimately connected with the existence of dynamic heterogeneities.
For example, Hurley and Harrowell used a model of fluctuating mobilities to account for the 
non-gaussian effects in a two-dimensional liquid \cite{softdisks}.
However, there are a few input parameters for this model,
which must be adjusted to explain the simulation results, e.g. the functional form of the mobility autocorrelation.
So, it is not clear to what extent the concept of fluctuating mobilities meets reality
and a deeper understanding is still
necessary to clarify the underlying physical mechanisms.

Again we analyse three-time correlations, now employing the information content
of the widths $\sigma_{zz,\|,\perp}$.
The important observation in figure \ref{fig:allall} is that the length of the first step of a tagged particle
has an influence on the mean size of its subsequent step, i.e. $\sizz$ and $\sigma_{\|,\perp}(\ra)$ grow with
$\za$ and $\ra$, respectively.
This can be understood in the
following way. A fast particle in the first time interval on average remains fast during the second one,
and so, by looking  at $p_{zz}(.|\za)$ or $p_{\|,\perp}(.|\ra)$ for large $\za$ or $\ra$,
we select the most mobile particles.
If we now calculate the degree, to which $\sigma_{zz}$ and $\sigma_{\|,\perp}$ grow with $\za$ and $\ra$, respectively,
we have a direct measure
to which extent the dynamics is ruled by heterogeneties. For this purpose we define the quantity
\begin{eqnarray}
a(t)\equiv\frac{\langle\sigma^4\rangle-\langle\sigma^2\rangle^2}{\langle\sigma^2\rangle^2},
\end{eqnarray}
where the brackets denote an average over the first step,
i.e. $\langle..\rangle\equiv\int \textrm{d}\ra..p(\ra)$ or
$\langle..\rangle\equiv\int \textrm{d}\za..p(\za)$.
Here, $a$,$\sigma$ and $p$ have to be
decorated with $zz$, $\|$ or $\perp$, depending on the probability function that is being analyzed.
(Note that a constant width $\sigma$ results in a vanishing $a(t)$.)

\begin{figure}[t]
\centerline{\includegraphics[clip, width=2in]{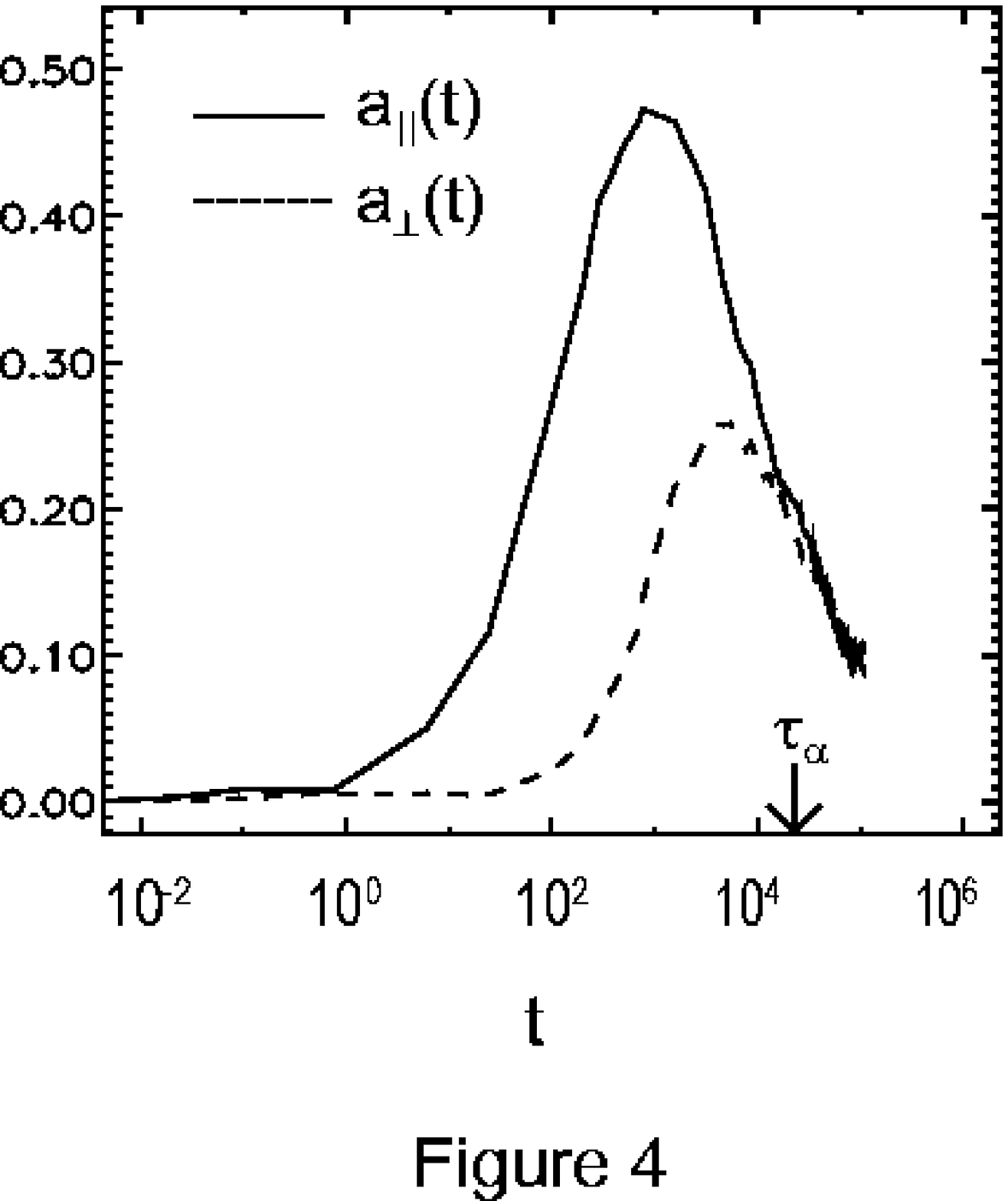}}
\renewcommand{\figurename}{Fig.}
{\footnotesize
The quantities $a_\|(t)$ and $a_\perp(t)$, as defined in the text, for the density $\phi=58\%$.
The difference in the maximum values illustrates the fact that heterogeneities at intermediate times
possess a directionality. For times greater than $\ta\approx2\cdot10^4$, both curves meet
each other, indicating the loss of anisotropy.
}
\label{fig:aparasenk}
\end{figure}

Figure \ref{fig:aparasenk} shows $a_\|(t)$ and $a_\perp (t)$ for the density $\phi=0.58$.
We see that the broadening with $\ra$, along the direction of the first step,
is greater than perpendicular to it, resulting in a larger value of $a_\|(t)$.
For longer times $t>\ta$, however, both curves coincide, indicating that there is no motional
anisotropy with respect to the previous step anymore.
Collective flow patterns, as observed for example by Donati et al. \cite{Donati}, could cause such a
directionality at intermediate times.

\begin{figure}[t]
\centerline{\includegraphics[clip, width=2in]{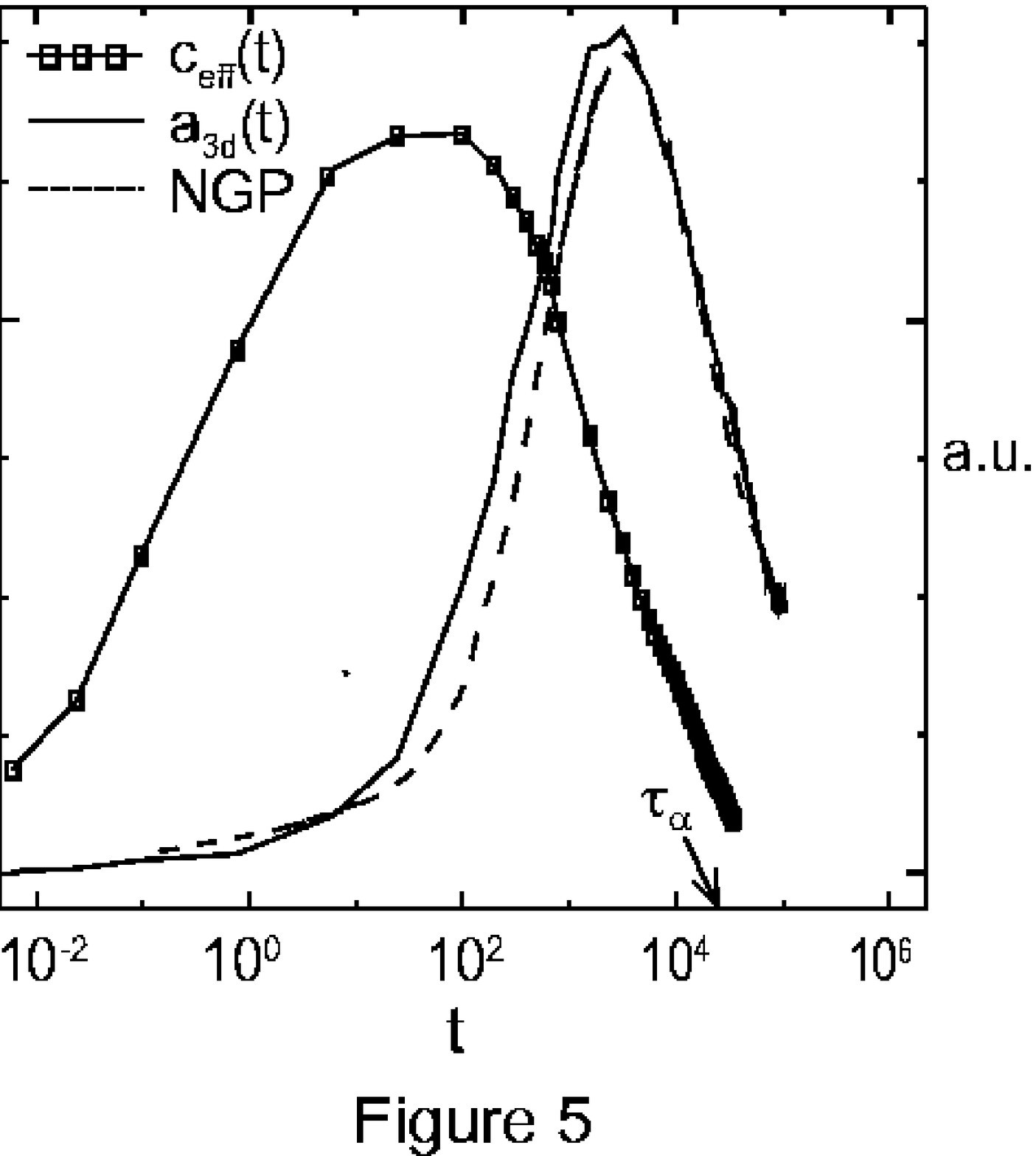}}
\renewcommand{\figurename}{Fig.}
{\footnotesize The averaged $a_{3d}\equiv(a_\|+2a_\perp)/3$, compared
with the NGP $\alpha_{2}$ and the effective slope
$\ceff$ for density $\phi=58\%$.
$a_{3d}$ and $\alpha_{2}$ show the same behaviour with time,
their maxima falling into the late $\beta$-region.
$\ceff$, in contrast, is large only for times $t<\ta$.
Note that the units are arbitrary.
}
\label{fig:aundNGP_poly}
\end{figure}

It is clear, that $a(t)\not=0$ should imply a nonvanishing NGP $\at$.
Fig. \ref{fig:aundNGP_poly} shows
the relation between $\alpha_2$ and the averaged $a_{3d}(t)\equiv\frac{a_\|+2a_\perp}3$.
This surprising similarity, concerning the time dependence of both curves, suggests a deep physical connection.
For a molecular liquid, this fact has already been reported by Qian et al. \cite{Jiang}, using another
quantity similar to $a_{3d}(t)$.
$\ceff$ is shown for a comparison of dynamical time regimes, reflecting the homogeneous contributions
to the complex relaxation process.

\section{Discussion}

As a major result of this work we were able to quantify the
homogeneous and the heterogeneous contributions to the
non-exponential relaxation. We found that  the anomalous diffusion
is mainly related to homogeneous contributions, explicitly
expressed by Eqs. (1) and (2), whereas the non-gaussian effects, displaying a
maximum for the crossover between $\beta$- and $\alpha$-relaxation,
are mainly related to the heterogeneous contributions. This
directly shows that for the system, studied in this work, the
non-gaussian effects, if at all, are only mildly related to jump
contributions. The maximum value of $\alpha_2$ is of the same order
as for the case of Lennard-Jones systems \cite{LJgross}. It remains the
interesting question whether also for the latter
$\alpha_2$ is mainly determined by heterogeneous contributions.

Relating the anomalous diffusion to homogeneous contributions,
as discussed above, is only valid for stationary processes.
As outlined, e.g., in Ref. \cite{Richert},
continuous time random walks, which by
definition are purely heterogeneous in nature, are also able to
generate anomalous diffusion behaviour. This, however, explicitly
requires non-stationary conditions.

We would like to mention that $a(t)$, defined in Eq.(3), only takes
into account heterogeneous contributions, related to the
$r_{01}$-dependence of $\sigma_{\|, \perp}(r_{01})$ and neglects
contributions resulting from the $r_{01}$-dependence of
$c_{\|}(r_{01})$. A simple model system with $\sigma_{\|,
\perp}(r_{01})= const$ and $c_{\|}(r_{01})\propto -r_{01}$ is an ensemble of
 diffusive
particles in identical harmonic potentials \cite{PRL}. The
dynamics is purely gaussian, i.e. $\alpha_2 \equiv 0$. Here the
heterogeneity is related to the fact that particles which, by
chance, are far away from the center of the potential will
experience a strong back-dragging force and will hence on average
move further in the near future. As discussed in \cite{Okun} this
effect can be also observed for the Rouse model of polymer
dynamics. Note that on a qualitative level this type of
heterogeneity is distinct from the type expressed by the
$r_{01}$-dependence of $\sigma_{\|,
\perp}(r_{01})$ for which the different mobilities are related
to different environments.

We gratefully acknowledge helpful discussions with S. B\"uchner and H.W. Spiess.


\begin{thebibliography}{99}
\bibitem{Boehmer}
B\"ohmer R, Chamberlin R V, Diezemann G, Geil B, Heuer A, Hinze G, Kuebler S C,
Richert R, Schiener B, Sillescu H, Spiess H W, Tracht U and Wilhelm M
1998 \emph{J. of Non-Cryst. Solids} \textbf{235-237} 1
\bibitem{Cichocki}
Cichocki B and Hinsen K 1990 \emph{Physica (Amsterdam)} \textbf{166A} 473
\bibitem{Donati}
Donati C, Douglas J F, Kob W, Plimpton S J, Poole P H and Glotzer S C 1998 \emph{Phys. Rev. Lett.} \textbf{80} 2338
\bibitem{Ediger}
Ediger M D, Angell C A and Nagel S R 1996 \emph{J. Phys. Chem.} \textbf{100} 13200 
\bibitem{Fuchs}
Fuchs M, G\"otze W and Mayr M R 1998 \emph{Phys. Rev.} E \textbf{58} 3384
\bibitem{softdisks}
Hurley M M and Harrowell P 1996 \emph{J. Chem. Phys.} \textbf{105} 10521
\bibitem{PRL}
Doliwa B and Heuer A 1998 \emph{Phys. Rev. Lett.} \textbf{80} 4915
\bibitem{LJgross}
Kob W and Andersen H C 1995 \emph{Phys. Rev.} E \textbf{51} 4626
\bibitem{vanMegen}
Megen van W and Pusey P N 1991 \emph{Phys. Rev.} A \textbf{43} 5429
\bibitem{Okun}
Heuer A and Okun K 1997 \emph{J. Chem. Phys.} \textbf{106} 6176
\bibitem{Jiang}
Qian J, Hentschke R and Heuer A 1998 \emph{J. Chem. Phys.}, submitted
\bibitem{Richert}
Richert R and Blumen A (eds.) 1994 \emph{Relaxational Processes} (Springer, Berlin)

\end{thebibliography}
\end{document}